\documentclass[manuscript]{aastex63}
\usepackage[T1]{fontenc}
\usepackage{wasysym}
\usepackage{xcolor}

\newcommand\mean[1]{\langle#1\rangle}
\newcommand\secref[1]{Section \ref{#1}}

\graphicspath{{./}{figures/}}

\usepackage{CJKutf8}
\AtBeginDvi{\input{zhwinfonts}} 
\newcommand{\Chi}[2]{%
  \csname CJK*\endcsname{UTF8}{zhsong}%
    \CJKchar{#1}{#2}%
  \csname endCJK*\endcsname
}
\DeclareUnicodeCharacter{9673}{\Chi{"96}{"73}}
\DeclareUnicodeCharacter{82F1}{\Chi{"82}{"F1}}
\DeclareUnicodeCharacter{540C}{\Chi{"54}{"0C}}
\DeclareUnicodeCharacter{738B}{\Chi{"73}{"8B}}
\DeclareUnicodeCharacter{7965}{\Chi{"79}{"65}}
\DeclareUnicodeCharacter{5B87}{\Chi{"5B}{"87}}

\received{2020 December 23}
\revised{2021 July 28}
\accepted{2021 August 9}
\submitjournal{the Planetary Science Journal}

\shorttitle{2013 VZ$_{70}$ and the Temporary Coorbitals of the Giant Planets}
\shortauthors{Alexandersen et al.}

\begin{document}
\title{\replaced{OSSOS XXI:\ Col3N10}{OSSOS.\ XXIII.\ 2013 VZ$_{70}$} and the Temporary Coorbitals of the Giant Planets}

\author[0000-0003-4143-8589]{Mike Alexandersen} 
\correspondingauthor{Mike Alexandersen}
\email{mike.alexandersen@alumni.ubc.ca}
\affiliation{Center for Astrophysics | Harvard \& Smithsonian, 60 Garden Street, Cambridge, MA 02138, USA}
\affiliation{Institute of Astronomy and Astrophysics, Academia Sinica; 11F of AS/NTU Astronomy-Mathematics Building, No. 1 Roosevelt Rd., Sec. 4, Taipei 10617, Taiwan}

\author[0000-0002-4439-1539]{Sarah Greenstreet}
\affiliation{B612 Asteroid Institute, 20 Sunnyside Ave, Suite 427, Mill Valley, CA 94941}
\affiliation{DIRAC Center, Department of Astronomy, University of Washington, 3910 15th Ave NE, Seattle, WA 98195}
\affiliation{Las Cumbres Observatory, 6740 Cortona Drive, Suite 102, Goleta, CA 93117, USA}
\affiliation{University of California, Santa Barbara, Santa Barbara, CA 93106, USA}

\author[0000-0002-0283-2260]{Brett J. Gladman}
\affiliation{Department of Physics and Astronomy, University of British Columbia, Vancouver, BC V6T 1Z1, Canada}

\author[0000-0003-3257-4490]{Michele T. Bannister}
\affiliation{School of Physical and Chemical Sciences -- Te Kura Mat\={u}, University of Canterbury, Private Bag 4800, Christchurch 8140, New Zealand}

\author[0000-0001-7244-6069]{Ying-Tung Chen (\Chi{"96}{"73}\Chi{"82}{"F1}\Chi{"54}{"0C})}
\affiliation{Institute of Astronomy and Astrophysics, Academia Sinica; 11F of AS/NTU Astronomy-Mathematics Building, No. 1 Roosevelt Rd., Sec. 4, Taipei 10617, Taiwan}

\author[0000-0001-8221-8406]{Stephen D. J. Gwyn}
\affiliation{Herzberg Astronomy and Astrophysics Research Centre, National Research Council of Canada, 5071 West Saanich Rd, Victoria, British Columbia V9E 2E7, Canada}

\author[0000-0001-7032-5255]{JJ Kavelaars}
\affiliation{Herzberg Astronomy and Astrophysics Research Centre, National Research Council of Canada, 5071 West Saanich Rd, Victoria, British Columbia V9E 2E7, Canada}
\affiliation{Department of Physics and Astronomy, University of Victoria, Elliott Building, 3800 Finnerty Rd, Victoria, BC V8P 5C2, Canada}

\author[0000-0003-0407-2266]{Jean-Marc Petit}
\affiliation{Institut UTINAM UMR6213, CNRS, Univ. Bourgogne Franche-Comt\'e, OSU Theta F25000 Besan\c{c}on, France}

\author[0000-0001-8736-236X]{Kathryn Volk}
\affiliation{Lunar and Planetary Laboratory, University of Arizona, 1629 E University Blvd, Tucson, AZ 85721, USA}

\author[0000-0003-4077-0985]{Matthew J. Lehner}
\affiliation{Institute of Astronomy and Astrophysics, Academia Sinica; 11F of AS/NTU Astronomy-Mathematics Building, No. 1 Roosevelt Rd., Sec. 4, Taipei 10617, Taiwan}
\affiliation{Department of Physics and Astronomy, University of Pennsylvania, 209 S. 33rd St., Philadelphia, PA 19104, USA}
\affiliation{Center for Astrophysics | Harvard \& Smithsonian, 60 Garden Street, Cambridge, MA 02138, USA}

\author[0000-0001-6491-1901]{Shiang-Yu Wang (\Chi{"73}{"8B}\Chi{"79}{"65}\Chi{"5B}{"87})}
\affiliation{Institute of Astronomy and Astrophysics, Academia Sinica; 11F of AS/NTU Astronomy-Mathematics Building, No. 1 Roosevelt Rd., Sec. 4, Taipei 10617, Taiwan}

\nocollaboration{11}


\begin{abstract}

We present the discovery of 2013 VZ$_{70}$, the first known horseshoe coorbital companion of Saturn. 
Observed by the Outer Solar System Origins Survey (OSSOS) for 4.5 years, the orbit of 2013 VZ$_{70}$ is determined to high precision, revealing that it currently is in `horseshoe' libration with the planet.
This coorbital motion will last at least thousands of years but ends $\sim10$~kyr from now; 2013 VZ$_{70}$ is thus another example of the already-known `transient coorbital' populations of the giant planets, with this being the first known prograde example for Saturn (temporary retrograde coorbitals are known for Jupiter and Saturn).
We present a theoretical steady state model of the scattering population of trans-Neptunian origin in the giant planet region (2--34 au), including the temporary coorbital populations of the four giant planets. 
We expose this model to observational biases using survey simulations in order to compare the model to the real detections made by a set of well-characterized outer Solar System surveys. 
While the observed number of coorbitals relative to the scattering population is higher than predicted, we show that the number of observed transient coorbitals of each giant planet relative to each other is consistent with a transneptunian source.
\footnote{This is a preprint. The nicely formatted, typo-free, final, open access, published version is available at \href{https://doi.org/10.3847/PSJ/ac1c6b}{https://doi.org/10.3847/PSJ/ac1c6b}}

\end{abstract}



\section{Introduction} \label{sec:intro}

Coorbital objects are found in the 1:1 mean-motion resonance with a planet. 
Resonance membership is determined by inspecting the evolution of the resonant angle $\phi_{11}=\lambda-\lambda_P$, where $\lambda=\Omega+\omega+M$ is the mean longitude, $P$ denotes the planet, $\Omega$ is the longitude of the ascending node, $\omega$ the argument of pericenter and $M$ the mean anomaly. 
The resonant angle $\phi_{11}$ must librate rather than circulate (ie. $\phi_{11}$ must occupy a bounded range) in order for an object to be considered to be in coorbital resonance.  
Like other $n:1$ resonances, the 1:1 mean-motion resonance includes multiple libration islands; objects in these islands are called leading Trojans (mean $\mean{\phi_{11}}=+60\degr$), trailing Trojans ($\mean{\phi_{11}}=300\degr=-60\degr$), quasi-satellites ($\mean{\phi_{11}}=0\degr$) or horseshoe coorbitals ($\mean{\phi_{11}}=180\degr$).
The motion of Trojans librate around one of the L4 or L5 Lagrangian points, while the path of horseshoe coorbitals encompass all of the L3, L4 and L5 Lagrangian points; quasi-satellites appear to orbit the planet (while not actually being bound to it). 
Quasi-satellites and horseshoe coorbitals are almost always unstable and thus temporary \citep[eg.][]{mikkola06, cuk12, jedicke18} with the exception of Saturn's moons Epimetheus and Janus, which are horseshoe coorbitals of each other \citep{fountainlarson78}.
\citet{greenstreet20} and \citet{li18} discuss the existence of high inclination ($i>90\degr$) objects temporarily trapped in a 1:-1 retrograde ``coorbital'' resonance with \added{Jupiter and }Saturn\added{, respectively}, although these are not coorbitals in the traditional sense described above; since they orbit the Sun in the opposite direction than \replaced{Saturn}{the planet}, retrograde coorbitals are not protected from close approaches with \replaced{Saturn}{the planet} the way that prograde coorbitals are, nor do the resonant island librations (ie. Trojan, horseshoe, quasi-satellite motion) behave in the traditional sense in the retrograde configuration. \explain{This last sentence only made mention of Saturn, but \citet{greenstreet20} is about Jovian retrograde coorbitals, so we have generalized the sentence.}

For many planets, the coorbital phase space is unstable due to perturbations from neighboring planets \citep[eg.][]{nesvornydones02, dvorak10}. 
\citet{innanenmikkola89} first suggested, at a time when only the Jovian Trojans were known, that populations of objects in stable 1:1 resonance with each of the other giant planets may exist; their analysis showed that the exact Lagrangian points are unstable for Saturn, but that Trojans farther from the resonance center (featuring larger libration amplitudes) could be stable for at least 10 Myr. 
These results were confirmed by \citet{holmanwisdom93}. 
Using longer timescales than previous studies, \citet{delabarre96} specifically studied the stability of Saturnian Trojans and found that Saturnian Trojans could only be long-term ($>428$ Myr) stable with very specific conditions: very small eccentricity (<0.028), $\phi_{11}$ libration amplitude greater than $80\degr$, $\omega$ libration about a point $45\degr$ ahead of Saturn's $\omega$, and constraints on the timing of the maximum eccentricity relative to the timing of Jupiter's maximum eccentricity, so that Jupiter and the Trojans do not approach close enough to dislodge the Trojan from Saturn's 1:1 resonance. 
\citet{nesvornydones02} showed that while Neptunian Trojans may have only been depleted by a factor of 2 over the age of the Solar System, the Saturnian Trojans would have been depleted by a factor of 100. 
Studying the cause of the instability of Saturnian Trojans, \citet{marzarischoll00} and \citet{hou14} found that the instability is caused by \replaced{mixed secular resonances, near-commensurability of libration frequencies and the vicinity of the Great Inequality (the near 5:2 resonance of Jupiter and Saturn)}{interactions between mean motion and secular resonances}. 
\added{\citet{huang19} investigated the stability of retrograde Saturnian coorbitals and found that they are always unstable due to an overlap with the $\nu_5$ and $\nu_6$ secular resonances.}
Given these destabilizing factors, causing any primordial population to have been mostly depleted\replaced{,}{ and} allowing only small niches to be long term stable, it is not surprising that no long-term stable Saturnian Trojans have been discovered to date.

Only Mars, Jupiter and Neptune have known populations of long-term ($>$Gyr) stable Trojans (which thus might be primordial) \citep{wolf1906, bowell90, levison97, marzari03, scholl05}.
\added{These long-term stable Trojan populations are important for understanding planet formation processes. As a few examples: 
\citet{polishook17} suggested that the Martian Trojans are likely to be impact ejecta from Mars, and used the mass of the current Trojan cloud to constrain how much Mars' orbit could have evolved during the phase of collisions. 
\citet{morbidelli05} showed that in order to reproduce the wide inclination-distribution of the Jovian Trojans, the Trojans must have been captured from an excited disk during a migration phase rather than having formed in place together with Jupiter. 
\citet{nesvorny13} demonstrated that a sudden displacement of Jupiter's semi-major axis, can explain the asymmetry seen between the L4 and L5 clouds and use the mass of the Jovian Trojan clouds to estimate the mass of the primordial planetesimal disk.
\citet{gomesnesvorny16} used the observed mass of Neptunian Trojans to infer that Neptune migrated slightly past its current location and then back, destabilizing the cloud, as we would otherwise observe a more massive cloud.
\citet{parker15} demonstrated that if Neptune's migration and eccentricity-damping was fast, the disk that it migrated into and captured Trojans from must already have been dynamically excited prior to Neptune's arrival in order to reproduced our observed orbital distribution.
}

\replaced{However}{While only three planets are known to have long-term stable Trojans}, scattering objects (scattering TNOs, Centaurs and even some objects originating in the asteroid belt\footnote{For the rest of this work, ``scattering objects'' will be considered synonymous with scattering TNOs and Centaurs of TNO origin, ignoring Centaurs originating from the asteroid belt, unless asteroidal origin is explicitly mentioned.}) can become temporary coorbitals, transiently captured into unstable resonance \citep{alexandersen13b, greenstreet20}. 
All Solar System planets except Mercury, Mars and Jupiter now have known populations of temporary coorbitals on prograde ($i<90\degr$) orbits \citep{wiegert98, mikkola04, karlsson04, hornerlykawka12, alexandersen13b, greenstreet20}.
Temporary ``sticking'' like this also occurs in other resonances \citep[eg. ][]{duncanlevison97, tsiganis00, alvarezcandal05, lykawka07, yu18, volk18}.
\added{While long-term stable Trojans inform us of conditions in the time of planet formation and migration, the temporarily captured coorbitals inform us about properties of the scattering population.
For example, \citet{alexandersen13b} confirmed the \citet{shankman13} finding that the size distribution of the scattering population must have a transition in order to explain the observed ratio of small nearby scattering object (including Uranian coorbitals) and larger more distant ones (including Neptunian coorbitals)}

\citet{hornerevans06} integrated the Centaurs known at the time, demonstrating that Centaurs do indeed get captured into temporary coorbital resonance with the giant planets, claiming that Jupiter should have by far the most temporary coorbitals, followed by Saturn and hardly any for Uranus and Neptune.
\citet{alexandersen13b} pointed out that using the known centaurs as the starting sample is biased towards having more objects nearer the Sun, and thus more captures for the inner giant planets, resulting in a disagreement with the sample of at-the-time known temporary coorbitals; 
they instead used a model that started with scattering TNOs that scatter inwards to become Centaurs and temporary coorbitals, to demonstrate that a TNO origin can explain the distribution of the temporary coorbitals of Neptune and Uranus. 

In this paper we describe the discovery of the first known Saturnian horseshoe coorbital, 2013 VZ$_{70}$, and demonstrate its temporary nature (\secref{sec:col3n10}). 
Furthermore we expand upon the analysis of \citet{alexandersen13b} to analyze the populations of temporary coorbitals of all four giant planets, in an attempt to demonstrate the likely origin of 2013 VZ$_{70}$ and similar objects. 
We use numerical integrations to construct a steady-state distribution model of the scattering Trans-Neptunian Objects (TNOs) and temporary coorbitals of the giant planets (\secref{sec:steadystate}). 
Lastly we use survey simulations, exposing our model to the survey biases of a well-understood set of surveys, in order to compare our theoretical predictions to real detections of this population (\secref{sec:surveysim}).

\section{Observations and orbit of 2013 VZ$_{70}$} \label{sec:col3n10}

\begin{figure}[htb]
\includegraphics[width=0.7\textwidth]{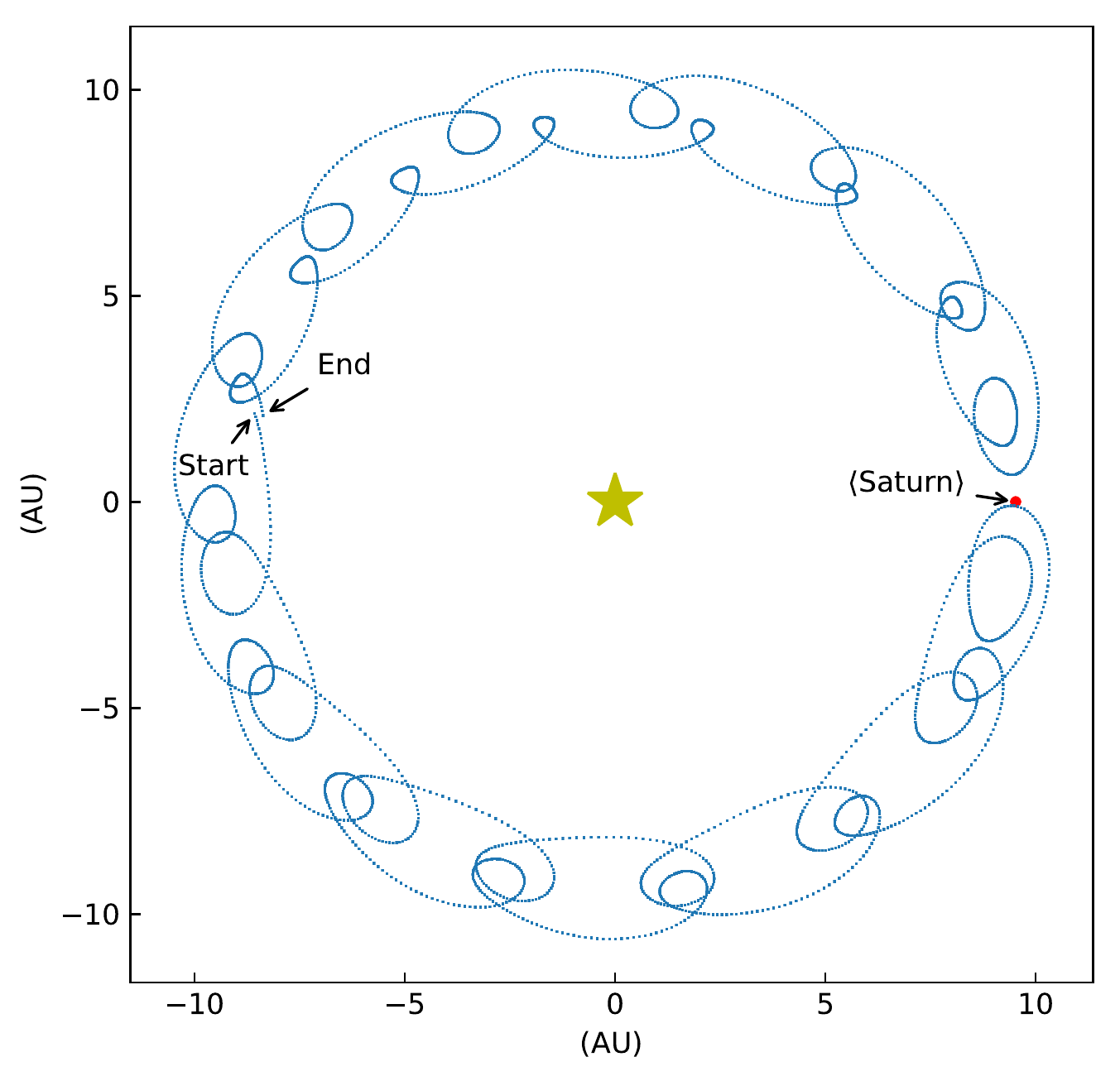}
\caption{\label{fig:corot}
The forward integrated motion of 2013 VZ$_{70}$ for 1 libration period (\replaced{$\sim1$~kyr}{$\sim890$~yr at this instance}), in the Saturnian mean-motion subtracted reference frame.
The blue dots shows the motion of 2013 VZ$_{70}$ relative to Saturn's mean motion (red dot).
The start and end points are marked.
The time between integration outputs (blue dots) is $\sim0.33$ years.
Note that when the object appears near Saturn in this planar, mean-motion subtracted projection, it is not actually close to the planet due to the vertical motion caused by the orbital inclination and the fact that Saturn's true location does oscillate around the marked mean location.
\added{The small cycles are caused by the eccentricity of the object's orbit, causing one little loop-and-shift motion for every orbit around the Sun.
Each local minimum in distance from the center of the plot corresponds to a perihelion passage, and each local maximum corresponds to an aphelion passage; the small loops occur when the object's distance is close to Saturn's mean heliocentric distance, while the large shifts occur when the object is at a \replaced{disctance}{distance} substantially different from Saturn, thus moving faster or slower around the Sun than Saturn does.
It is thus clear from this figure that the horseshoe libration period at this time is roughly 30 orbital periods, although the libration period does vary slightly while always remaining near $\sim1$~kyr.}
}
\end{figure}

2013 VZ$_{70}$ was discovered by the Outer Solar System Origins Survey \citep[OSSOS, ][]{bannister16,bannister18} in images taken on 2013 November 1 using the MegaCam wide-field imager \citep{boulade03} on the Canada-France-Hawai'i Telescope (CFHT). 
The object was subsequently measured in 37 tracking observations from 2013 August 09 to 2018 January 18 \citep[for the full list of astrometric measurements, see MPEC 2021-Q55, ][]{mpec_col3n10}.  
With 4.5 years of high-accuracy astrometry, the orbit is very well known, being $a=9.1838\pm0.0002$~AU, $e=0.097145\pm0.000011$, $i=12\degr.04110\pm0\degr.00006$, $\Omega=215\degr.22021\pm0\degr.00008$, $\omega=245\degr.754\pm0\degr.006$, $M=291\degr.425\pm0\degr.006$ for epoch = JDT 2456514.0. 
Here $a$, $e$, $i$ are the barycentric semi-major axis, eccentricity and inclination, respectively; the uncertainties are calculated from a covariance matrix using the orbit-fitting software \textit{Find\_Orb} \citep{gray11a}\added{and the JPL DE430 planetary ephemerides \citep{folkner14}}. 
This orbit is very close to that of Saturn\replaced{(although $\sim180\degr$ away)}{, although the two bodies are seperated by $\sim180\degr$ on the sky}.
From dynamical integrations we found that 2013 VZ$_{70}$ is in fact in the 1:1 mean motion resonance with Saturn, in a horseshoe configuration (see \autoref{fig:corot}).
However, the best fit clone only remains resonant for about 11~kyr before leaving the resonance and re-joining the scattering population. 

\begin{figure}[htb]
\includegraphics[width=1.0\textwidth]{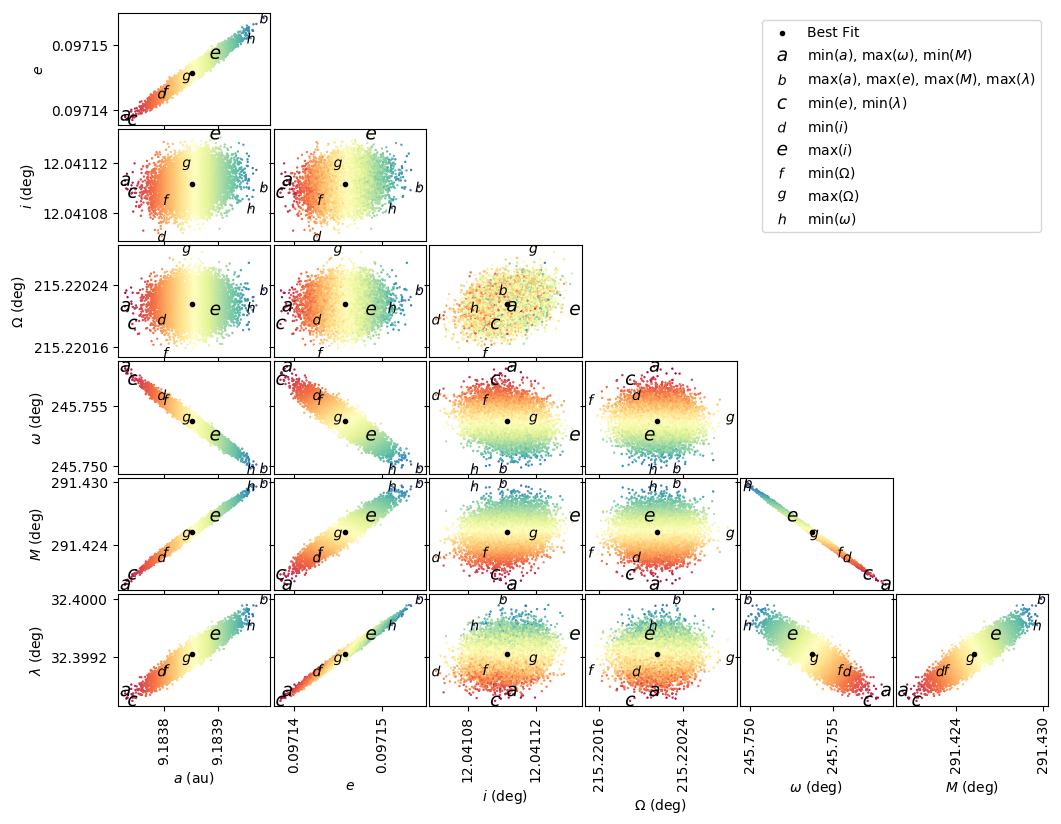}
\caption{\label{fig:mcmc_clones}
Orbital elements of 10,000 fits to resampled astrometry. 
Each clone was generated using \emph{Find\_Orb}'s MCMC feature, using a Gaussian noise equal to the mean residual of the best fit orbit ($0.\arcsec146$). 
The clones are color coded by semi-major axis (which is also the x-axis of the left most row) to give an additional rough indicator of its correlation with the other orbital elements. 
The best fit orbit is marked with a black dot, and orbits with either the smallest or largest value of one of the parameters are marked with a letter ((a-h, not to be confused with any orbital elements). 
This figure demonstrates how the uncertainties on the different parameters are related; most parameters are strongly coupled, while $i$ and $\Omega$ have only weak or no coupling with other parameters. 
}
\end{figure}

\begin{figure}[htb]
\includegraphics[width=1.0\textwidth]{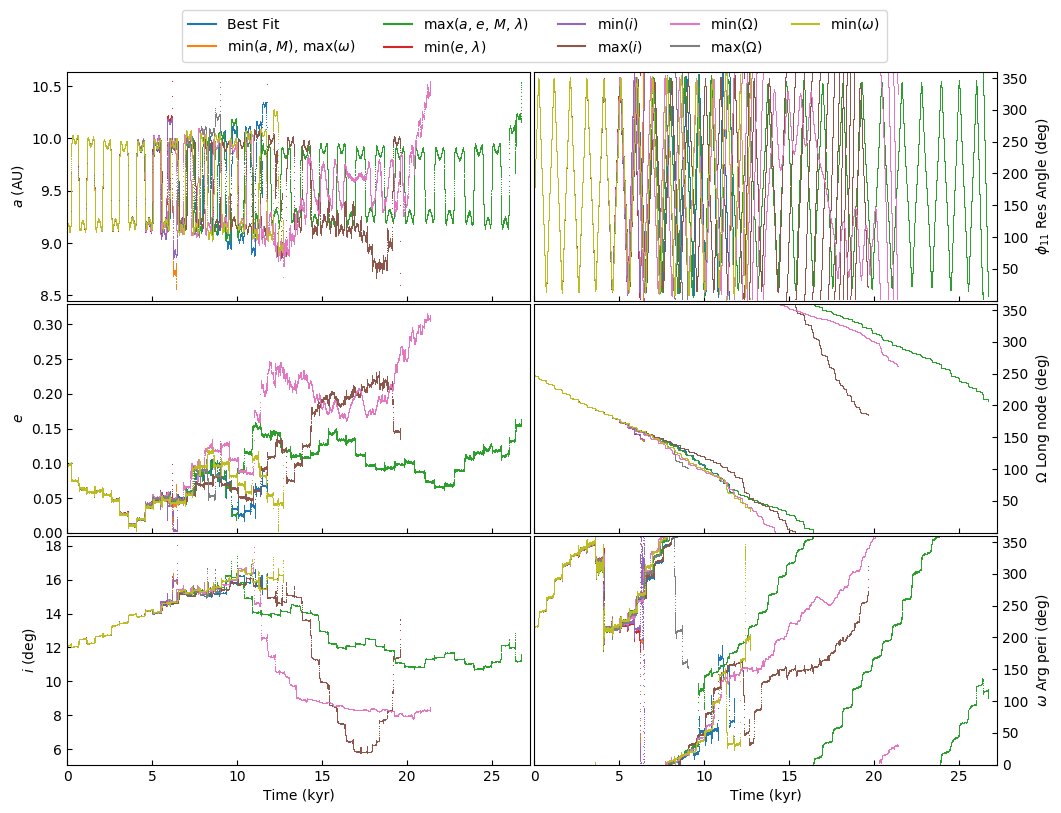}
\caption{\label{fig:el_vs_time}
Future evolution of 2013 VZ$_{70}$, the temporary Saturnian horseshoe coorbital. 
The nine clones marked on \autoref{fig:mcmc_clones} were integrated (the best fit orbit plus eight extremal clones).
For clarity, clone trajectories have only been drawn until the semi-major axis of the clone deviates from Saturn's by more than 1 au for the first time. 
The top right panel has been expanded in \autoref{fig:resangle_vs_time} to better show the clones' interactions with the resonance. 
The "stair step" patterns occurs at the time when \replaced{$\phi_11$}{$\phi_{11}$} is close to $0\degr/360\degr$, which is the time the coorbital is closest to the planet; the close approach causes the switch from a slightly-larger-than-the-planet's semi-major axis to a slightly-smaller-than-the-planet's (and vice versa) that ensures that the planet/coorbital never overtake each other. 
The close approaches also imparts small changes in the other orbital elements, seen as the "stair step" pattern.
}
\end{figure}

\begin{figure}[htb]
\includegraphics[width=1.0\textwidth]{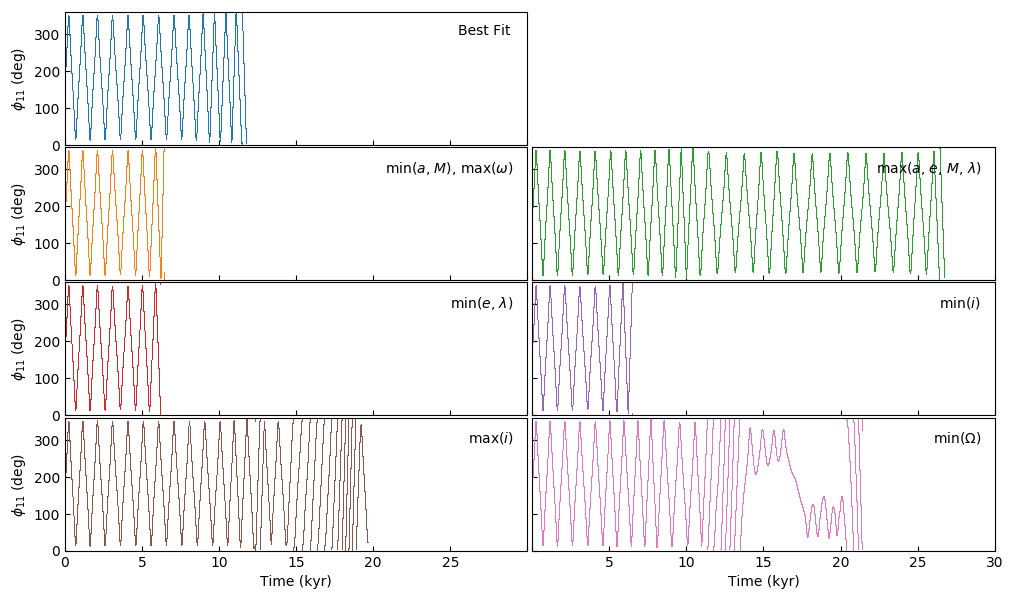}
\caption{\label{fig:resangle_vs_time}
Future evolution of the resonant angle $\phi_{11}$ (with respect to Saturn) for each of the 2013 VZ$_{70}$ clones shown in Figures \ref{fig:mcmc_clones} and \ref{fig:el_vs_time}.
The information here is identical to that in the top-right subplot of \autoref{fig:el_vs_time}, but split up for clarity. 
As in \autoref{fig:el_vs_time}, clone trajectories have only been drawn until the semi-major axis of the clone deviates from Saturn's by more than 1 au for the first time. 
It is clear that 2013 VZ$_{70}$ is currently in the Saturnian coorbital resonance, but will escape from this state in 6--26 kyr. 
Note that after leaving the resonance, the minimum $\Omega$ clone gets recaptured and librates a few times in each Trojan island before becoming re-ejected. 
}
\end{figure}

We investigated whether the best fit orbit could be near a stability boundary by generating orbit clones from appropriate resampling of our astrometry, allowing us to test whether any orbit consistent with the astrometry featured long-term stability.
Each clone was produced by resampling all the astrometry (using a normal distribution with standard deviation equal to the mean residual of the best fit, $0.\arcsec146$) and fitting a new orbit.
This process was repeated 10,000 times using \textit{Find\_Orb}.
The distribution of orbits generated by this process explicitly shows how the uncertainty of some of the orbital parameters are strongly coupled, as can be seen in \autoref{fig:mcmc_clones}.
From these 10,000 clones, we identified the most extreme orbits (largest and smallest value of each parameter) and integrated these 8 clones (labelled in \autoref{fig:mcmc_clones}) as well as the best fit orbit. 
These dynamical integrations were done using \textit{Rebound}\citep{reinliu12} with the \textit{WHFast}\citep{wisdomholman91, kinoshita91, reintamayo15} symplectic integrator. 
The eight major planets and Pluto\footnote{Pluto was primarily included as a test to ensure that the system was set up correctly, not because we expect the mass of Pluto to have any influence on the outcome of the integration. However, since Pluto's mass is known, there was no reason to not include it. Pluto was confirmed to be resonating in the 3:2 resonance with Neptune in our integrations, as expected.} were included as massive perturbers and an integration step size of 5\% of Mercury's initial orbital period ($\simeq0.012$ yr $\simeq4.39$ days) was used, while output was saved approximately 3 times per year.
As can be seen in Figures \ref{fig:el_vs_time} and \ref{fig:resangle_vs_time}, the future evolution of all of the clones involve an initial period in coorbital resonance, but all clones leave the resonance between 6~kyr and 26~kyr from now. 
The large range of resonance exit times is due to the highly chaotic nature of the orbit. 
We used a second set of numerical integrations, where clones were displaced infinitesimally ($10^{-13}$--$10^{-12}$ AU, or 1.5--15 cm) relative to \replaced{the best fit clone}{each of the above clones}, to estimate the object's current timescale for chaotic divergence (the Lyapunov time scale); we found this to be \replaced{$\sim16$}{$410\pm60$} yr\added{\footnote{A Jupyter notebook demonstrating how the Lyapunov time scale was calculated is available at DOI: \href{https://doi.org/10.11570/21.0008}{10.11570/21.0008}.}}.
2013 VZ$_{70}$ is thus definitely in coorbital resonance now, but the chaotic nature of the orbit means that the duration of this temporary resonance capture will likely not be constrained further, even with additional observations.

\section{Deriving the steady state orbital distribution} \label{sec:steadystate}

We proceed to investigate the potential origin of temporary coorbitals like 2013 VZ$_{70}$. 
We model the source of the giant planet temporary coorbitals and investigate their detectability in characterised surveys. 
We produced a steady-state distribution of scattering objects in the $a<34$~au region from orbital integrations similar to those used in \citet{alexandersen13b}; the details of those integrations can be found in the supplementary material of that paper. 
We primarily outline the deviations from those used in the previous paper below.

To perform the dynamical integrations, we used the N-body code SWIFT-RMVS4 (provided by Hal Levison, based on the original SWIFT \citep{levisonduncan94}) with a base time step of 25 days and an output interval of 50 years for the orbital elements of the planets and any particle which at the moment had $a<34$~au.
The gravitational influences of the four giant planets and the Sun were included.
The system starts with 8500 particles, derived from the 34~au $< a <$ 200~au scattering portion of the \citet{kaib11} model of the outer Solar System.
Particles were removed from the simulation when they hit a planet, went outside 2000~AU or inside 2~AU from the Sun (since they would either interact with the terrestrial planets that are absent in our simulations or would rapidly be removed from the Solar System by Jupiter), or the final integration time of 1~Gyr was reached.
\added{Since 1~Gyr is substantially longer than the dynamical lifetime of Centaurs and scattering TNOs, we thoroughly sample the $a<$34~au phase space, despite the limited number of initial particles. 
The output is combined along the time-axis to produce a distribution of approximately $300$ million sets of orbital elements.}
As in \citet{alexandersen13b} we confirm that the distribution in the first 100~Myr is similar enough to the distribution in the following 900~Myr (because the \replaced{a<}{$a<$}34~au region is populated very quickly despite starting off empty) that we can treat the distribution as a whole as being in steady state. 
\added{We also ran a similar simulation with particles drawn from the modified version of the \citet{kaib11} model also used in \citet{alexandersen13b} and \citet{shankman13}; this modified version was generated with an assumption of the primordial planetesimal disk being more dynamically excited. 
As in \citet{alexandersen13b}, we find that using the standard \citet{kaib11} model and the modified version as our starting condition makes little quantitative difference on the end results. 
For the rest of this work we will therefore only be referring to the results from the simulations using the standard \citet{kaib11} model.}

The method for determining coorbital behavior in the particle histories is also very similar to that used in \citet{alexandersen13b}, with some small modifications.
To diagnose whether particles are coorbital, the orbital histories (at 50 year output intervals) were scanned using a running window 30~kyr long for Uranus/Neptune and 5~kyr long for Jupiter/Saturn;
this window size was chosen to be several times longer than the typical Trojan libration period at the given planet ($\sim1$~kyr for Saturn as seen in \autoref{fig:resangle_vs_time}). 
A particle was classified as a coorbital if, within the running window, both its average semimajor axis was less than 0.2~AU from the average semimajor axis of a given planet and no individual semimajor axis value deviated more than $R_H$ from that of the planet. 
Here $R_H$ is the planet's Hill sphere radius \citep{murraydermott99}, where $R_H = 0.35$~ AU for Jupiter, $R_H = 0.44$~AU for Saturn, $R_H = 0.47$~AU for Uranus, and $R_H = 0.77$~AU for Neptune. 
Further determination of which resonant island a coorbital is librating in was made identically to the method used in \citet{alexandersen13b}.

Our results are in good agreement with those for Uranus and Neptune in \citet{alexandersen13b}. \autoref{tab:steady} contains the fraction of the steady-state population in coorbital motion with each of the giant planets at any given time, as well as the distribution of coorbitals between horseshoe, Trojan and quasi-satellite orbits.
The coorbital fraction for Uranus and Neptune are slightly higher than in \citet{alexandersen13b}, despite very similar methodology; however, these results agree within their expected accuracy.
The capture fraction decreases from Neptune through to Jupiter (with almost 4000 times fewer coorbitals than Neptune), although this is unsurpring given that the source of the scattering objects is beyond Neptune and that the dynamical timescales (orbital period, libration timescale) are longer farther from the Sun. 
An interesting result is that while Neptune's and Uranus' coorbitals are roughly equally distributed between horseshoe and Trojan coorbitals, Saturn seems to very preferentially capture scattering objects into horseshoe orbits, and Jupiter has a much larger fraction of quasi-satellites than any of the other planets. 
\autoref{tab:steady} also shows the mean, median and maximum duration of a capture in coorbital resonance with the planets, as well as the mean, median and maximum number of captures experienced by particles with at least one episode of coorbital motion with a given planet. 
The mean and median coorbital lifetimes and number of captures for Uranus and Neptune are also within a factor of two of those in \citet{alexandersen13b}, which we thus adopt as the uncertainty. 
Note that the captures into coorbital motion with Saturn and Jupiter are typically significantly shorter than for Uranus and Neptune, although if the lifetimes are represented \added{in }units of orbital periods rather than years, the Jovian captures actually have the second longest lifetimes, after Uranus. 
\added{While the different coorbital resonances no doubt experience different interactions with secular resonances and experience different perturbations from neighbouring planets, it is noteworthy that the median number of orbital periods for coorbital captures for all four planets are within a factor of 2.5 of each other.}
However, going Neptune to Jupiter, particles are increasingly unlikely to have multiple captures, presumably due to the increasing ability of the planet to scatter the object to large semi-major axis; this results in particles on average spending both more total time and more total orbital periods in coorbital motion with Uranus and Neptune than Saturn and Jupiter. 

\begin{deluxetable}{l|D|CCC|DDD|C|DDD}[tbp]
\tabletypesize{\footnotesize}
\tablecaption{\label{tab:steady} 
Steady state fractions of the $a<34$~AU, $q>2$~AU scattering objects that are in temporary co-orbital resonance with the giant planets. 
For reference, ``Horseshoe'' coorbitals librate about $\phi_{11}=180\degr$, ``Trojan'' librate about $\phi_{11}=60\degr$ and $300\degr$, and ``Quasi-satellites'' librate about $\phi_{11}=0\degr$.
Also listed are the mean, median and maximum duration of such captures seen in our simulations, and the median lifetime divided by the orbital period of the associated planet. 
Lastly the mean, median and maximum number of captures experienced by a particle that is trapped by the planet at least once.
}
\tablehead{
\colhead{Planet} & \twocolhead{Coorbitals} & \colhead{Horseshoe} & \colhead{Trojan} & \colhead{Quasi-satellite} & \multicolumn6c{Lifetime (kyr)}                              & \colhead{Median lifetime}   & \multicolumn6c{Number of traps}  \\
\colhead{}       & \twocolhead{$\%$ of scattering} & \multicolumn3c{$\%$ of planet's coorbitals}                & \twocolhead{Mean} & \twocolhead{Median} & \twocolhead{Max}  & \colhead{(orbital periods)} & \twocolhead{Mean} & \twocolhead{Median} & \twocolhead{Max}}
\decimals                                                                                                                                                                                                                                   
\startdata                                                                                                                                                                                                                                  
Jupiter          & 0.00093                         & 33                  & 21               & 46                &  11               &  7.1                &     26            & 600                         &  1                &  1                  &  1 \\
Saturn           & 0.022                           & 85                  & 12               &  3                &  19               & 10                  &    630            & 340                         &  2                &  1                  &  6 \\
Uranus           & 0.65                            & 56                  & 37               &  7                & 129               & 59                  & 16,000            & 700                         &  4                &  2                  & 30 \\
Neptune          & 3.6                             & 48                  & 40               & 12                &  83               & 46                  &  3,300            & 280                         & 10                &  5                  & 85 \\
\enddata
\end{deluxetable}

\section{Comparing theory and observations} \label{sec:surveysim}

In order to compare our dynamical model to the real detections, we run the model through the OSSOS Survey Simulator \citep{bannister18, lawler18b}. 
The survey simulator generates one object at a time (with an orbit drawn from the dynamical model and an $H$-magnitude drawn from a parametric model discussed later) and assesses whether the object would have been discovered by the input surveys. 
We used all of the characterized surveys with sufficient characterization available for use in the simulator: the Canada-France Ecliptic Plane Survey \citep[CFEPS, ][]{petit11}, the CFEPS High-Latitude extension \citep[HiLat]{petit15}, the \citet{alexandersen16} survey, and OSSOS \citep{bannister18}.
These surveys combined will be referred to as OSSOS++. 

\subsection{Orbital distribution}

For our survey simulations, it is preferable to have orbital distribution functions rather than an orbital distribution composed of a fixed number of discrete particles.
This allows for the simulator to be run for as long as necessary, without producing duplicate identical particles. 
We have set up independent distributions for the coorbitals and the scattering objects, as described below, both inspired by the distribution seen in the integrations discussed in \secref{sec:steadystate}.
\added{Our model files and scripts for use with the OSSOS Survey Simulator \citep{lawler18b} are provided at DOI: \href{https://doi.org/10.11570/21.0008}{10.11570/21.0008} for anybody curious to use this model distribution.}\explain{This DOI will be published when the manuscript is accepted for publication.}

\subsubsection{Scattering objects}\label{sec:scatteringmodel}

We use the output from the \secref{sec:steadystate} integrations, taking every particle's orbit at every time step and binning them using bin sizes of $0.5$~au, $0.02$ and $2\degr.0$ in $a$, $e$ and $i$ space. 
The survey simulator reads this binned table, randomly selects a bin weighted by the number of particles that went into the bin, and then randomly assigns $a$, $e$ and $i$ from a uniform distribution within the bin. $\Omega$, $\omega$ and $M$ are all assigned randomly from a uniform distribution from $0\degr$ to $360\degr$, since the orientations of scattering objects' orbits are random. 
This process allows us to draw essentially infinite unique particles that follow a distribution consistent with the steady-state distribution from \secref{sec:steadystate}.

\subsubsection{Coorbital objects} \label{sec:coorbmodel}

We cannot simply bin the coorbital distributions as we did for the scattering distribution. 
The numbers of coorbitals in the \secref{sec:steadystate} integrations are low (particularly for Jupiter), and the number of dimensions we would need to bin is higher since the resonant angle $\phi_{11}$ is also important for the coorbital distribution. 
Instead, we opted to use parametric distributions, fitted to the distributions seen in \secref{sec:steadystate}. 

In this simplified parametric model, the semi-major axis of the coorbital is always set equal to that of the planet, since the few tenths of au variability do not influence detectability by sky surveys as much as the details of the eccentricity and inclination distribution. 
The eccentricity is modelled with a normal distribution, centred at 0 with a width $w_e$, multiplied by $\sin^2(e)$, truncated to [$e_{min}$, $e_{max}$]\footnote{In the end, $e_{min}\approx0.0$ was always best, but this was not required.}:
\begin{equation}
    f(e|e_{min}\leq e\leq e_{max}) = \frac{\sin^2(e)}{w_e\sqrt{2\pi}}\exp\left(\frac{-e^2}{2w_e^2}\right)
\end{equation}
This functional form has little physical motivation and was merely chosen as it in the end provides a good fit to the distribution seen in our integrations. 
The inclination is modelled as a normal distribution, with centre at $0\degr$ and a width $w_i$, multiplied by $\sin(i)$, truncated at $i_{max}$:
\begin{equation}
    f(i|0\degr\leq i\leq i_{max}) = \frac{\sin(i)}{w_i\sqrt{2\pi}}\exp\left(\frac{-i^2}{2w_i^2}\right)
\end{equation}
This is simply a Normal distribution modified to account for the spherical coordinate system. 
Lastly, $\Omega$ and $M$ are chosen randomly from a uniform distribution [$0\degr$, $360\degr$) while $\omega$ is calculated from $\phi_{11}$, the value of which depends on the type of coorbital. 
The different types of coorbitals are generated using the ratios in \autoref{tab:steady}. 
The details of the selection of a $\phi_{11}$ value is similar to that used in \citet{alexandersen13b}, accounting for a distribution of libration amplitudes and the fact that the centre of libration is offset away from $60\degr$/$300\degr$ for Trojans with large libration amplitudes. 
\deleted{We provide all of our model files and scripts in the supplementary files for anybody curious to use this model distribution.}
The values of $w_e$, $w_i$, $e_{min}$, $e_{max}$, and $i_{max}$ used in this work for each planet's coorbital population are shown in \autoref{tab:orbitparams}.

\begin{deluxetable}{lDDDDD}[tbp]
\tablecaption{\label{tab:orbitparams} 
Orbital parameters used for each planet in the parametric model described in Section \ref{sec:coorbmodel}.
}
\tablehead{
\colhead{Planet} & \twocolhead{$w_e$} & \twocolhead{$e_{min}$} & \twocolhead{$e_{max}$} & \twocolhead{$w_i$ ($\degr$)} & \twocolhead{$i_{max}$ ($\degr$)}}
\decimals                                                                                                        
\startdata                                                                                                       
Jupiter          & 0.188    & 0.0 & 0.523    & 16.1     & 40.6     \\
Saturn           & 0.127    & 0.0 & 0.707    & 15.6     & 89.6     \\
Uranus           & 0.134    & 0.0 & 0.998    & 19.4     & 51.3     \\
Neptune          & 0.123    & 0.0 & 0.974    & 18.1     & 80.9     \\
\enddata
\end{deluxetable}


\subsection{Absolute magnitude distributions}

The Solar System absolute magnitude ($H$) distribution of the TNOs is not well constrained for objects fainter than about $H_r\approx8.0$, although it is clear that there is a transition from a steep to shallower slope somewhere in $7.5\leq H_r\leq9$ \citep{sheppardtrujillo10b, shankman13, fraser14, alexandersen16, lawler18}. 
The scattering objects provide a clue to the small-end distribution, as many of these reach distances closer to the Sun, allowing us to more easily detect smaller objects. 
\citet{lawler18} carefully analysed the size distribution of the scattering objects in OSSOS++; since our sample is a subset of their sample (we only use objects with $a<34$ au), we will directly apply the two magnitude distributions favored by \citet{lawler18}: a divot (with $\alpha_{b}=0.9$, $\alpha_{f}=0.5$, $H_{b}=8.3$ and $c=3$) and a knee (with $\alpha_{b}=0.9$, $\alpha_{f}=0.4$, $H_{b}=7.7$ and $c=1$).
Here $\alpha_{b}$ and $\alpha_{f}$ are the exponent of the exponential magnitude distribution on the bright and faint (respectively) side of a transition that happens at the break magnitude $H_b$; $c$ denotes the contrast factor of the population immediately on each side of the break, such that $c=1$ is a knee and $c>1$ is a divot. 
For further details on this parameterization, see \citet{shankman13} and \citet{, lawler18}.

\subsection{Population estimate} \label{sec:popest}

\startlongtable
\begin{deluxetable}{clCDcDDD@{${}\pm{}$}LD@{${}\pm{}$}LD}
\tabletypesize{\scriptsize}
\tablecaption{\label{tab:sample}
Details of the sample (29 objects) used in this work.
MPC name denotes the Minor Planet Center designation for the TNO, while O++ name is the internal designation used within the OSSOS++ surveys. 
Cls is the classification of the object, where coorbitals of Saturn, Uranus and Neptune are indicated with the initial of the planet (S, U, N), with subscripted H, 4 or 5 for horseshoe coorbitals, leading Trojans and trailing Trojans, respectively; finally C indicates a non-coorbital Centaur/scattering object. 
Mag is the magnitude at discovery in the filter F, while $H$ is the absolute magnitude in that same filter. 
The J2000 barycentric distance, semi-major axis, eccentricity and inclination are shown in $d$, $a$, $e$, $i$, respectively; for both $d$ and $i$ the uncertainty is 1 on the last digit or smaller, and has therefore been omitted. The elements for 2013 VZ$_{70}$ were calculated using \textit{Find\_Orb} \citep{gray11a} and the JPL DE430 planetary ephemerides \citep{folkner14}, while elements for all other objects were taken from \citet{bannister18}. 
}
\tablehead{
\colhead{MPC}  & \colhead{O++} & \colhead{Cls} & \twocolhead{Mag} & \colhead{F} & \twocolhead{$H$} & \twocolhead{$d$} & \multicolumn3c{$a$}  & \multicolumn3c{$e$} & \twocolhead{$i$}     \\
\colhead{name} & \colhead{name}    & \colhead{}      & \twocolhead{}    & \colhead{}       & \twocolhead{}    & \twocolhead{(au)}    & \multicolumn3c{(au)} & \multicolumn3c{}    & \twocolhead{$\degr$}
}
\decimals
\startdata
2013 VZ$_{70}$         & Col3N10 & S_H & 23.28 & r & 13.75 &  8.891 &  9.1838 & 0.0002 & 0.097145 & 0.000011 & 12.041	\\ 
2015 KJ$_{172}$        & o5m02   & C   & 24.31 & r & 14.68 &  9.180 & 10.8412 & 0.0018 & 0.47436  & 0.00012  & 11.403	\\ 
2015 GY$_{53}$         & o5p001  & C   & 24.05 & r & 13.40 & 12.029 & 12.0487 & 0.0011 & 0.0828   & 0.0003   & 24.112	\\ 
2015 KH$_{172}$        & o5m01   & C   & 23.55 & r & 14.92 &  7.434 & 16.896  & 0.004  & 0.68003  & 0.00010  & 9.083	\\ 
(523790) 2015 HP$_{9}$ & o5p003  & C   & 21.39 & r & 10.15 & 13.563 & 18.146  & 0.003  & 0.2699   & 0.0003   & 3.070	\\ 
2011 QF$_{99}$         & mal01   & U_4 & 22.57 & r &  9.56 & 20.296 & 19.092  & 0.003  & 0.1769   & 0.0004   & 10.811	\\ 
2013 UC$_{17}$         & o3l02   & C   & 23.86 & r & 11.42 & 17.045 & 19.3278 & 0.0008 & 0.12702  & 0.00004  & 32.476	\\ 
2015 RE$_{277}$        & o5t01   & C   & 24.02 & r & 16.13 &  6.018 & 20.4545 & 0.0012 & 0.766535 & 0.000014 & 1.621	\\ 
2015 RH$_{277}$        & o5s04   & C   & 24.51 & r & 13.11 & 13.441 & 20.916  & 0.008  & 0.5083   & 0.0003   & 10.109	\\ 
2015 GB$_{54}$         & o5p004  & C   & 23.92 & r & 12.68 & 13.563 & 20.993  & 0.007  & 0.4205   & 0.0003   & 1.628	\\ 
2015 RF$_{277}$        & o5t02   & C   & 24.91 & r & 14.51 & 10.616 & 21.692  & 0.004  & 0.51931  & 0.00013  & 0.927	\\ 
2015 RV$_{245}$        & o5s05   & C   & 23.21 & r & 10.10 & 19.884 & 21.981  & 0.010  & 0.4793   & 0.0003   & 15.389	\\ 
2013 JC$_{64}$         & o3o01   & C   & 23.39 & r & 11.95 & 13.774 & 22.145  & 0.002  & 0.37858  & 0.00006  & 32.021	\\ 
2015 GA$_{54}$         & o5p005  & C   & 24.34 & r & 10.67 & 23.500 & 22.236  & 0.007  & 0.2582   & 0.0006   & 11.402	\\ 
2014 UJ$_{225}$        & o4h01   & C   & 22.74 & r & 10.29 & 17.756 & 23.196  & 0.009  & 0.3779   & 0.0004   & 21.319	\\ 
2013 UU$_{17}$         & o3l03   & C   & 24.07 & r &  9.93 & 25.336 & 25.87   & 0.04   & 0.249    & 0.003    & 8.515	\\ 
2015 RD$_{277}$        & o5t03   & C   & 23.27 & r & 10.48 & 18.515 & 25.9676 & 0.0014 & 0.28801  & 0.00004  & 18.849	\\ 
2015 RK$_{277}$        & o5s01   & C   & 23.36 & r & 15.29 &  6.237 & 26.9108 & 0.0012 & 0.802736 & 0.000009 & 9.533	\\ 
2014 UG$_{229}$        & o4h02   & C   & 24.33 & r & 11.47 & 19.526 & 27.955  & 0.005  & 0.44082  & 0.00011  & 12.242	\\ 
2015 VF$_{164}$        & o5d001  & C   & 23.93 & r & 12.74 & 13.286 & 28.273  & 0.005  & 0.54257  & 0.00013  & 5.729	\\ 
2015 VE$_{164}$        & o5c001  & C   & 23.72 & r & 11.75 & 15.857 & 28.529  & 0.007  & 0.45711  & 0.00019  & 36.539	\\ 
2012 UW$_{177}$        & mah01   & N_4 & 24.20 & r & 10.61 & 22.432 & 30.072  & 0.003  & 0.25912  & 0.00016  & 53.886	\\ 
2004 KV$_{18}$         & L4k09   & N_5 & 23.64 & g &  9.33 & 26.634 & 30.192  & 0.003  & 0.1852   & 0.0003   & 13.586	\\ 
2015 RU$_{245}$        & o5t04   & C   & 22.99 & r &  9.32 & 22.722 & 30.989  & 0.007  & 0.2898   & 0.0003   & 13.747	\\ 
2015 GV$_{55}$         & o5p019  & C   & 22.94 & r &  7.55 & 34.605 & 31.375  & 0.011  & 0.3026   & 0.0005   & 28.287	\\ 
2008 AU$_{138}$        & HL8a1   & C   & 22.93 & r &  6.29 & 44.517 & 32.393  & 0.002  & 0.37440  & 0.00009  & 42.826	\\ 
2015 KS$_{174}$        & o5m04   & C   & 24.38 & r & 10.19 & 26.018 & 32.489  & 0.005  & 0.2254   & 0.0002   & 7.026	\\ 
2004 MW$_{8}$          & L4m01   & C   & 23.75 & g &  8.75 & 31.360 & 33.467  & 0.004  & 0.33272  & 0.00008  & 8.205	\\ 
2015 VZ$_{167}$        & o5c002  & C   & 23.74 & r & 11.18 & 17.958 & 33.557  & 0.005  & 0.52485  & 0.00009  & 15.414	\\ 
\enddata
\end{deluxetable}

We predict a population estimate for the scattering objects with $a<34$~au of trans-Neptunian origin based on our model and the real detections. 
The OSSOS++ surveys discovered a total of 29 scattering objects with $a<34$~au (including 4 temporary coorbitals), listed in \autoref{tab:sample}. 
For the purposses of this work, 2013 VZ$_{70}$ is included in this sample, despite its uncharacterized status, as discussed in \replaced{\autoref{sec:col3n10}}{\autoref{sec:caveat}}.
We thus ran the survey simulation with our scattering model (see \secref{sec:scatteringmodel}) as input until it detected 29 objects, recorded how many objects had been drawn from the model, and repeated 1000 times to measure the uncertainty for the population estimate. 
For the divot and knee $H$ distributions respectively, we predict the existence of $(2.1\pm0.2)\times10^7$ and $(4.9\pm1.0)\times10^6$ scattering TNOs with $a<34$~au and $H_r<19$. 
Given the size and orbit distribution, most of these are small objects beyond $30$~au and thus far beyond the detectability both of the surveys we consider here and similar-depth future surveys like the upcoming Legacy Survey of Space and Time (LSST) on the Vera Rubin Observatory. 

\subsection{Expected versus detected numbers}

Using the population estimate of the $a<34$ au scattering objects as measured in \secref{sec:popest}, we predict the number of temporary coorbitals of TNO origin that OSSOS++ should have detected. 
This is done by running the survey simulator for each planet's coorbital population separately (using the coorbital model defined in \secref{sec:coorbmodel}), inputting a fixed number of coorbital particles (equal to the total scattering object population estimate found in \secref{sec:popest} multiplied by the coorbital fraction for the given planet as found in \secref{sec:steadystate}) and recording the number of detections, repeating 1000 times to sample the distribution. 
We find that for both the divot and knee distribution and for each planet, the most common (expected) value of temporary coorbital detections is zero. 
However, the probabilities of getting zero detected temporary coorbitals for Saturn, Uranus and Neptune are $84\%$, $74\%$ and $71\%$ (divot) or $91\%$, $73\%$ and $59\%$ (knee). 
The probability of getting zero detections for all three planets in these surveys is thus less than $50\%$. 
In other words, more often than not, we would expect OSSOS++ to detect at least one giant planet coorbital beyond Jupiter (the case of Jupiter is discussed in the next paragraph).
From the distribution of simulated detections, we find that the detection of four coorbitals (as in the real surveys) is unlikely, at a probability of $0.8\%$, but not completely implausible. 
We expand on this below.  

For Jupiter, the chance of zero detections is $>99.99\%$ due to the rate of motion cuts imposed on/by the moving object detection algorithms of the OSSOS++ surveys; only the most eccentric Jovian coorbitals would have been detectable at aphelion (and only in a few fields).
It is thus entirely reasonable that OSSOS++ found no Jovian coorbitals, neither temporary nor long term stable; these surveys were simply not sensitive to objects at those distances. 
We note that there is one known temporary retrograde ($i>90\degr$) coorbital of Jupiter, 2015 BZ$_{509}$ (514107) Ka`epaoka`awela \citep{wiegert17}, whose origin is, according to \replaced{\citep{greenstreet20}}{\citet{greenstreet20}}, most likely the main asteroid belt and not the trans-neptunian/scattering object population. 
Our simulations produce no retrograde coorbitals of any of the planets, supporting that Ka`epaoka`awela likely originates from the asteroid belt and not the transneptunian region. 

Our survey simulations predict a number ratio of detected Jovian, Saturnian, Uranian and Neptunian coorbitals of 0:1:2:2 (J:S:U:N, where the mean number of detections have been scaled such that the value for Saturn is 1, then rounded). 
The ratio of real detections is 0:1:1:2 (2013 VZ$_{70}$, 2011 QF$_{99}$, 2012 UW$_{177}$ and 2004 KV$_{18}$), so the ratio of detected temporary coorbitals of each of the giant planets is in good agreement with predictions.
However, the survey simulations predict that only $3\%$ of the detected $a<34$~au scattering objects should be coorbitals, whereas the four real coorbitals make up $14\%$ of detections (4 of 29);
the observed fraction of the $a<34$~au scattering objects that are in temporary coorbital resonance is thus $\sim5$ times higher than expected. 
Before the OSSOS survey, which was by far the most sensitive survey of the ensemble and discovered over $80\%$ of the OSSOS++ TNOs and scattering objects, \replaced{$50\%$ were coorbital (3 of 6)}{$60\%$ were coorbital (3 of 5)}, so it would appear that the initially high fraction of coorbitals detected in the earlier surveys in our set was a fluke, and that the ratio is approaching the theoretical value predicted above as the observed sample increases.  \explain{We seem to have miscounted the non-OSSOS objects in \autoref{tab:sample}.}
We thus do not feel it justified to hypothesize additional sources for the temporary coorbital population at this time.
While we cannot rule out that the population of temporary coorbitals, particularly for Jupiter, is supplemented from other sources such as the asteroid belt and primordial Jovian Trojans, \citet{greenstreet20} finds that for Jovian temporary coorbitals, the asteroid belt is only the dominant source for retrograde ($i>90\degr$) coorbitals, which they estimate comprise $\ll 1\%$ of the\added{ temporary} coorbital population. 
It is unlikely that the asteroid belt is a dominant source for the outer planets if it isn't for Jupiter. 
The contribution of the asteroid belt to the steady state temporary coorbital distribution of the giant planets is thus insignificant, and we are likely not missing any important source population in producing our population/detection estimates. 

\subsection{Caveat}\label{sec:caveat}

While \replaced{Col3N10's orbit}{the orbit of 2013 VZ$_{70}$} was well determined by the OSSOS observations, it is not part of the characterized OSSOS dataset. 
2013 VZ$_{70}$ was discovered in images taken in a ``failed'' observing sequence from 2013B (failed due to poor image quality and the sequence not being completed), which was thus not used for the characterised (ie. well understood) part of the OSSOS survey. 
This failed sequence, which should have been 30 high-quality images of 10 fields (half of the OSSOS ``H'' block), only obtained low-quality (limiting $m_r\approx23.5$) images of 6 fields. 
A TNO search of these images was conducted (discovering 2013 VZ$_{70}$) to facilitate follow-up observations (color and light curve measurements), but this shallow search was never characterised due to the expectation that everything would be rediscovered in an eventual high-quality discovery sequence.  
A high-quality observing sequence of the full set of H-block fields was successfully observed in 2014B, with limiting magnitude $m_r=24.67$, which was used for the characterised search. 
However, as a year had passed, 2013 VZ$_{70}$ had already left the field due to its large rate of motion; unlike all other objects discovered in the failed 2013B sequence, 2013 VZ$_{70}$ was thus not re-discovered in the characterised discovery images. 
As such, 2013 VZ$_{70}$ is not part of the characterised sample of the survey, as that sample only includes objects discovered in specific images on specific nights through a carefully characterised process.
However, because the failed discovery sequence points at the same area of the sky as parts of the characterised survey and it is a small minority of the total observed fields, it would make hardly any difference on the discovery biases whether these particular images are included in the characterization or not.
From our simulations in \secref{sec:surveysim} we can see that only about $8\%$ of simulated detections of theoretical Saturnian Coorbitals were discovered in the OSSOS H block; this block is thus not in a crucial location for discovering Saturnian coorbitals in any way. 
The fact that the only Saturnian coorbital to have been discovered in OSSOS++ was among the very small minority of those surveys' total discoveries that were not characterised thus appears to be a low-probability event. 
We can therefore treat 2013 VZ$_{70}$ as effectively being part of the characterised survey for the purposes of this work, with the warning that this approach should not be used for other non-characterised objects from these surveys; most other objects are non-characterised for other reasons, mostly for being fainter than the well-measured part of the detection efficiency function.
\added{That being said, ignoring 2013 VZ$_{70}$ from the sample on grounds of being uncharacterized would brings the coorbital to total scattering ratio down to $11\%$ (3 out of 28), closer to the $3\%$ predicted in the previous section.}

\section{Conclusions} \label{sec:conclusions}

2013 VZ$_{70}$ is the first known temporary Saturnian horseshoe coorbital, remaining resonant for $6$--$26$~kyr; it likely originates in the trans-Neptunian region. 
Our simulations show that all the giant planets should have temporary coorbitals of TNO origin, although Jupiter has approximately a factor of 4000 fewer than Neptune; the duration of the coorbital captures are significantly more short-lived for Saturn and Jupiter than for Uranus and Neptune. 
Our simulations show that Neptune's and Uranus' coorbitals should be roughly equally distributed between horseshoe and Trojan coorbitals, Saturn very preferentially captures scattering objects into horseshoe orbits, and Jupiter should have a much larger fraction of its temporary coorbitals be quasi-satellites than any of the other planets. 
Accounting for observing biases in a set of well-characterized surveys (CFEPS \citep{petit11}, HiLat \citep{petit15}, the \citet{alexandersen16} survey, and OSSOS \citep{bannister18}) we find that the fraction of $a<34$~au scattering objects that are in temporary coorbital motion is higher in the real observations ($13.7\%$) than in simulated observations ($2.9\%$).
However, for the distribution of the temporary coorbitals among the giant planets, we find that our predictions ($\sim$ 0:1:2:2 for J:S:U:N) are consistent with the observations (0:1:1:2).

\acknowledgments

This work is based on TNO discoveries obtained with MegaPrime/MegaCam, a joint project of the Canada France Hawaii Telescope (CFHT) and CEA/DAPNIA, at CFHT which is operated by the National Research Council (NRC) of Canada, the Institute National des Sciences de l'Universe of the Centre National de la Recherche Scientifique (CNRS) of France, and the University of Hawaii. 
A portion of the access to the CFHT was made possible by the Academia Sinica Institute of Astronomy and Astrophysics, Taiwan. 
This research used the facilities of the Canadian Astronomy Data Centre operated by the National Research Council of Canada with the support of the Canadian Space Agency.

The authors wish to recognize and acknowledge the very significant cultural role and reverence that the summit of Maunakea has always had within the indigenous Hawaiian community. 
We are most fortunate to have the opportunity to conduct observations from this mountain.
We would also like to acknowledge the maintenance, cleaning, administrative and support staff at academic and telescope facilities, whose labor maintains the spaces where astrophysical inquiry can flourish.

The authors wish to thank Hanno Rein for useful discussions and help regarding how to estimate the Lyapunov timescale through numerical integrations.
The authors also wish to thank Jeremy Wood for the suggestion of looking at the capture durations in terms of orbital periods rather than only absolute time. 

SG acknowledges support from the Asteroid Institute, a program of B612, 20 Sunnyside Ave, Suite 427, Mill Valley, CA 94941. Major funding for the Asteroid Institute was generously provided by the W.K. Bowes Jr. Foundation and Steve Jurvetson. Research support is also provided from Founding and Asteroid Circle members K. Algeri-Wong, B. Anders, R. Armstrong, G. Baehr, The Barringer Crater Company, B. Burton, D. Carlson, S. Cerf, V. Cerf, Y. Chapman, J. Chervenak, D. Corrigan, E. Corrigan, A. Denton, E. Dyson, A. Eustace, S. Galitsky, L. \& A. Fritz, E. Gillum, L. Girand, Glaser Progress Foundation, D. Glasgow, A. Gleckler, J. Grimm, S. Grimm, G. Gruener, V. K. Hsu \& Sons Foundation Ltd., J. Huang, J. D. Jameson, J. Jameson, M. Jonsson Family Foundation, D. Kaiser, K. Kelley, S. Krausz, V. Lašas, J. Leszczenski, D. Liddle, S. Mak, G.McAdoo, S. McGregor, J. Mercer, M. Mullenweg, D. Murphy, P. Norvig, S. Pishevar, R. Quindlen, N. Ramsey, P. Rawls Family Fund, R. Rothrock, E. Sahakian, R. Schweickart, A. Slater, Tito’s Handmade Vodka, T. Trueman, F. B. Vaughn, R. C. Vaughn, B. Wheeler, Y. Wong, M. Wyndowe, and nine anonymous donors.
SG acknowledges the support from the University of Washington College of Arts and Sciences, Department of Astronomy, and the DIRAC Institute. The DIRAC Institute is supported through generous gifts from the Charles and Lisa Simonyi Fund for Arts and Sciences and the Washington Research Foundation.
This work was supported in part by NASA NEOO grant NNX14AM98G to LCOGT/Las Cumbres Observatory.

KV acknowledges support from NASA (grants NNX15AH59G and 80NSSC19K0785) and NSF (grant AST-1824869).

This work was supported by the Programme National de Plantologie (PNP) of CNRS-INSU co-funded by CNES. 

This research has made use of NASA's Astrophysics Data System Bibliographic Services. 
This research made use of SciPy \citep{jones01}, NumPy \citep{vanderwalt11}, matplotlib \citep[a Python library for publication quality graphics\added{, }][]{hunter07}.

\vspace{5mm}
\facilities{CFHT(MegaCam)}

\software{
Python \citep{vanrossumdeboer91},
Matplotlib \citep{hunter07},
NumPy \citep{vanderwalt11},
SciPy \citep{jones01},
Find\_Orb \citep{gray11a},
Rebound \citep{reinliu12},
fit\_radec \& abg\_to\_aei \citep{bernsteinkhushalani00},
SWIFT-RMVS4 \citep{levisonduncan94},
Jupyter notebook \citep{jupyter_notebooks}
}


\bibliography{ossos_coorbitals}


\listofchanges

\end{document}